\documentclass[aps,physrev,twocolumn,reprint,amsmath,amssymb]{revtex4-2}

\newcommand{\td}{\mathrm{d}}
\newcommand{\ti}{\mathrm{i}}
\newcommand{\te}{\mathrm{e}}

\usepackage{bm}
\usepackage{graphicx}
\usepackage{hyperref}

\begin{document}
 \title{Theory of generalized Josephson effects}

\author{Aron J. Beekman}
\email{aron@phys-h.keio.ac.jp}
\affiliation{Department of Physics, and Research and Education Center for Natural Sciences, Keio University, 3-14-1 Hiyoshi, Kohoku-ku, Yokohama 223-8522, Japan}
\date {\today}
\begin{abstract}
 The DC Josephson effect is the flow of supercurrent across a weak link between two superconductors with a different value of their order parameter, the phase. We generalize this notion to any kind of spontaneous continuous symmetry breaking. The quantity that flows between the two systems is identified as the Noether current associated with the broken symmetry. The AC Josephson effect is identified as the oscillations due to the energy difference between the two systems caused by an imposed asymmetric chemical potential. As an example of novel physics, a Josephson effect is predicted between two crystalline solids, potentially measurable as a force periodic in the separation distance.
\end{abstract}

\maketitle

The prediction by Josephson~\cite{Josephson62} of a DC current flowing between two superconductors separated by a small distance, with zero voltage bias, revealed two important concepts. First, the order parameter can rightfully be regarded as a quantum field with an equation of motion, that does not vanish abruptly at the edge of a sample but must fall off in a continuous fashion. Second, it settled the debate around broken symmetry in superconductors~\cite{Anderson64,Josephson74}. It may then be expected that a similar effect occurs in any system with spontaneously broken symmetry, since the only prerequisite seems to be a coupling between 
two systems whose order parameters take different values. Indeed, such generalizations have been explored. Of course the phenomenon shows up in helium-4 superfluidity because the broken symmetry is essentially the same as in superconductors, barring the coupling to gauge fields~\cite{Anderson66}. Josephson tunneling is also found in superfluid helium-3, which breaks a non-Abelian symmetry~\cite{Ambegaokar76}. The Josephson effect has further been generalized to $SO(5)$-symmetry proposed for high-$T_\mathrm{c}$ superconductivity~\cite{Zhang97,Demler98}. A spin current flowing between two ferromagnets~\cite{Nogueira04,Lee03,Chen14,Ruckriegel17} or antiferromagnets~\cite{Chasse10,Moor12,Liu16} is another manifestation of the Josephson effect.

Nevertheless, the Josephson effect as an intrinsic property of spontaneous symmetry breaking has received little attention, apart from Esposito {\em et al.}~\cite{Esposito07}. They elegantly describe the Josephson effect of a system with internal symmetry group $O(N)$ as the appearance of pseudo-Goldstone modes due to the explicit but weak breaking of the doubled symmetry group $O(N) \times O(N)$ that governs the two uncoupled systems, to the diagonal subgroup $O(N)$ that leaves the total charge invariant while the relative charge is broken. %Their restrictions to internal symmetries, and to type-A Goldstone modes, are presently shown to be unnecessary. 
This method has been applied to spinor Bose--Einstein condensates~\cite{Qi09}. 

\begin{figure}[b]
 \includegraphics[width=.85\columnwidth]{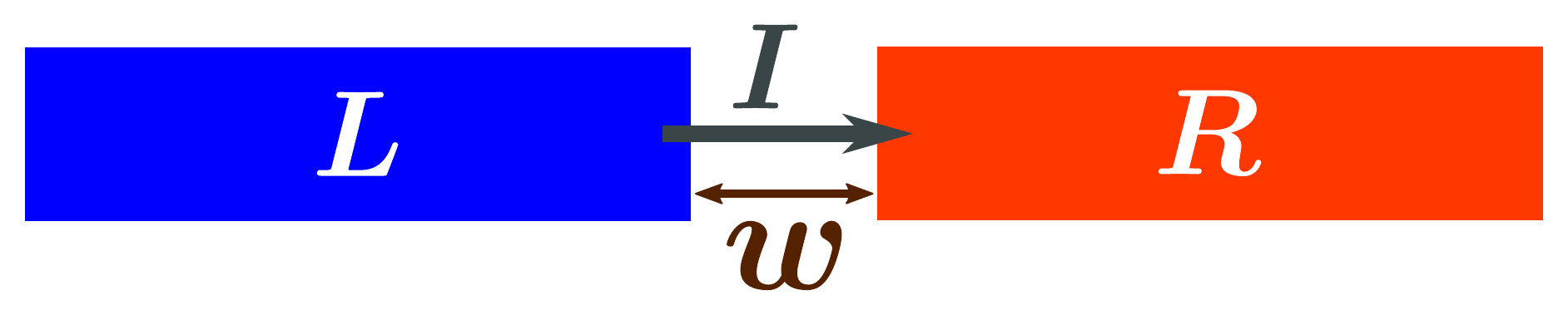}
 \caption{Typical Josephson setup. Two systems left and right are separated by a small distance $w$ so that their order parameters are coupled. Josephson current $I$ can flow from one system to the other.}
 \label{fig:setup}
\end{figure}

We derive the Josephson effect purely in terms of symmetry transformations of the order parameter operator $\mathcal{O}$. Suppose there are two systems on the left ($\mathrm{L}$) and right ($\mathrm{R}$) in states $\lvert \psi_\mathrm{L} \rangle$, $\lvert \psi_\mathrm{R} \rangle$ which break some symmetry by the formation of uniform order parameter expectation values $\mathbf{m}_\mathrm{L} = \langle \psi_\mathrm{L} \rvert \mathcal{O}_\mathrm{L} \lvert \psi_\mathrm{L} \rangle$ and $\mathbf{m}_\mathrm{R} =  \langle \psi_\mathrm{R} \rvert \mathcal{O}_\mathrm{R} \lvert \psi_\mathrm{R} \rangle$ (see Fig.~\ref{fig:setup}). The only assumption is that the order parameters of the two systems couple to each other in a way that favors a uniform configuration. This assumption is very plausible if one considers the two systems as parts of one large system that therefore prefers to break the symmetry uniformly. Explicitly, the coupling Hamiltonian takes the form:
\begin{equation}\label{eq:Josephson coupling}
 H_K = -\frac{K}{2} \left( \mathcal{O}^\dagger_\mathrm{L} \mathcal{O}_\mathrm{R} +  \mathcal{O}^\dagger_\mathrm{R} \mathcal{O}_\mathrm{L} \right).
\end{equation}
This term must be compensated by $H_{K'} = K \lvert \mathcal{O}_\mathrm{L} \rvert\lvert \mathcal{O}_\mathcal{R} \rvert$ so that the energy vanishes when the order parameters on left and right are the same. 
The coupling parameter $K>0$ is small with respect to other energy scales.

Without the coupling Eq.~\eqref{eq:Josephson coupling}, the Hamiltonian $H_0$ is invariant under a continuous symmetry group $G$. We can consider the two systems independently, with symmetry generators $Q_\mathrm{L}^a$ and $Q_\mathrm{R}^a$, where $a$ runs over the dimensions of the Lie algebra of $G$. A symmetry generator is the volume integral of a Noether charge density: $Q^a_{\mathrm{L},\mathrm{R}} = \int_{V \in \mathrm{L},\mathrm{R}} j_t^a(x,t)$, and the Noether currents are locally conserved $\partial_t j_t^a(x,t) + \nabla \cdot \mathbf{j}^a(x,t) =0$. We assume that the symmetry is broken down spontaneously to a subgroup $H \subset G$, due to formation of the order parameters $\mathbf{m}_\mathrm{L}$, $\mathbf{m}_\mathrm{R}$. These order parameters transform under some representation of $G$, where the generators $Q^a$ are represented by Hermitian matrices $T^a_{\phantom{a}ij}$, so that a general transformation generated by $Q^a_\mathrm{L}$ is parametrized as
\begin{equation}\label{eq:order parameter rotation}
 m_{\mathrm{L},i} \to \te^{\ti \alpha T^a_{ij}} m_{\mathrm{L},j},\qquad \alpha \in \mathbb{R},
\end{equation}
and the same for the right system. For convenience we have chosen $\mathbf{m}_\mathrm{L}$, $\mathbf{m}_\mathrm{R}$ to transform under a vector representation, but this is not necessary for anything that follows. This transformation is equivalent to
\begin{equation}\label{eq:order parameter commutator}
 \langle [Q^a_\mathrm{L},\mathcal{O}_\mathrm{L}] \rangle = \langle T^a \mathcal{O}_\mathrm{L} \rangle = T^a_{ij} m_{\mathrm{L},j},
\end{equation}
where by abuse of notation we have also used $T^a$ as the linear transformation acting on the operator $\mathcal{O}$. By Hermitian conjugation $[Q^a,\mathcal{O}^\dagger_\mathrm{L}] = - \mathcal{O}^\dagger_\mathrm{L} T^{a\dagger} = - \mathcal{O}^\dagger_\mathrm{L} T^a$ because $Q^a$ and $T^a$ are Hermitian.

The space of values that the order parameters $\mathbf{m}_\mathrm{L}$ and $\mathbf{m}_\mathrm{R}$ can take is isomorphic to the quotient space $G/H$, and Eq.~\eqref{eq:order parameter rotation} can be considered as a rotation in that space. We are interested in the situation where $\mathbf{m}_\mathrm{L}$ and $\mathbf{m}_\mathrm{R}$ take different values in $G/H$. All the symmetry groups we consider are transitive, which implies that we can always choose the generators $T^a$ in such a way that a single generator $T^\mathrm{r}$ connects the right system to the left:
\begin{align}\label{eq:order parameter phase difference}
 \mathbf{m}_\mathrm{L} &= m_\mathrm{L} \mathbf{m}_0, &
 \mathbf{m}_\mathrm{R} &= m_\mathrm{R}\te^{\ti \alpha T^\mathrm{r}}\mathbf{m}_0.
\end{align}
Here $\mathbf{m}_0$ is some unit vector, and $m_\mathrm{L}$, $m_\mathrm{R}$ are the magnitudes of the order parameters.

%%%%%%%%%%%%%%%%%%%%%%%%%%%%%

\textit{DC Josephson effect.}
Turning on the coupling Eq.~\eqref{eq:Josephson coupling}, the Hamiltonian no longer commutes with $Q^a_\mathrm{L}$ and $Q^a_\mathrm{R}$ separately, although it does commute with $Q^a_\mathrm{L} +  Q^a_\mathrm{R}$. We can now calculate the expectation value with respect to the zero-coupling ground state  $\lvert \psi_\mathrm{L} \rangle \otimes \lvert \psi_\mathrm{R} \rangle$ of the time derivative of the Noether charge of one system, say the left, using the Heisenberg equation of motion:
\begin{align}
  \langle \partial_t Q^a_\mathrm{L} \rangle 
 &= -\frac{1}{\ti\hbar} \langle [H ,Q^a_\mathrm{L}] \rangle \notag\\
 &= \frac{K}{2\ti \hbar} \langle \left( [\mathcal{O}^\dagger_\mathrm{L},Q^a_\mathrm{L}] \mathcal{O}_\mathrm{R} +  \mathcal{O}^\dagger_\mathrm{R} [\mathcal{O}_\mathrm{L},Q^a_\mathrm{L}] \right) \rangle \notag\\
 &= \frac{K}{2\ti \hbar} \langle \left( \mathcal{O}^\dagger_\mathrm{L} T^a \mathcal{O}_\mathrm{R} - \mathcal{O}^\dagger_\mathrm{R} T^a \mathcal{O}_\mathrm{L} \right)\rangle \notag\\
 &= \frac{K}{2\ti \hbar} \left( \mathbf{m}_\mathrm{L}^\dagger T^a \mathbf{m}_\mathrm{R} - \mathbf{m}_\mathrm{R}^\dagger T^a \mathbf{m}_\mathrm{L} \right).  
\end{align}
Here we have used the fact that $\mathcal{O}_\mathrm{L}$ and $\mathcal{O}_\mathrm{R}$ commute and act on different parts of Hilbert space, so we can take expectation values of their products trivially. The last line immediately shows that there is no current if $\mathbf{m}_\mathrm{R}/m_\mathrm{R} = \mathbf{m}_\mathrm{L} / m_\mathrm{L}$. But if the order parameter of the right system is rotated with respect to the left and takes the form of Eq.~\eqref{eq:order parameter phase difference} we find the general form of the DC Josephson effect:
\begin{equation}
  \langle \partial_t Q^a_\mathrm{L} \rangle  
 = \frac{K}{2\ti\hbar}m_\mathrm{L}m_\mathrm{R} \mathbf{m}_0^\dagger \left(T^a \te^{-\ti \alpha T^\mathrm{r}} - \te^{\ti \alpha T^\mathrm{r}} T^a \right) \mathbf{m}_0.
 \label{eq:generalized Josephson effect}
\end{equation}
In all cases of interest the right-hand side is only non-zero when $a = \mathrm{r}$, for which the equation simplifies to
\begin{align}
 \langle \partial_t Q^\mathrm{r}_\mathrm{L} \rangle  
 &= -\frac{K}{\hbar}m_\mathrm{L}m_\mathrm{R} \mathbf{m}_0^\dagger T^\mathrm{r} \sin (\alpha T^\mathrm{r}) \mathbf{m}_0.
 \label{eq:generalized Josephson effect commute}
\end{align}
Eqs.~\eqref{eq:generalized Josephson effect} and \eqref{eq:generalized Josephson effect commute} are the main result of this work. 

We now consider its physical implications. First of all, the current is seen to be 
the flow of Noether charge associated with the broken generator $Q^\mathrm{r}$ that connects the two systems. This can be made more explicit, by writing the left-hand side of Eq.~\eqref{eq:generalized Josephson effect} as
\begin{align}\label{eq:boundary current}
 \langle \partial_t Q^a_\mathrm{L} \rangle 
 &= \int_V \td^3 x\; \langle \partial_t j^a_{\mathrm{L},t} \rangle
 =- \int_V \td^3 x\; \langle \nabla \cdot  \mathbf{j}^a_{\mathrm{L}} \rangle  \notag\\
 &= - \oint_{\partial V} \td \mathbf{S} \cdot \langle \mathbf{j}^a_\mathrm{L} \rangle = -I_\mathrm{boundary}.
\end{align}
The change of Noether charge $\langle Q^a_\mathrm{L} \rangle$ is caused by flow of Noether current $\langle \mathbf{j}^a_\mathrm{L} \rangle$ through the boundary $\partial V$, and the sign of $I$ denoting flow out of the left system (see Fig.~\ref{fig:setup}). This connects with the aforementioned work by Esposito {\em et al.}~\cite{Esposito07} since the broken Noether charge density excites Goldstone modes. It also demonstrates that the spin Josephson current between two ferromagnets and that between two antiferromagnets is the same, because they break the same symmetry, even though their order parameters are different. Consequently there is no difference in DC Josephson effect between systems with type-A or type-B Goldstone modes.

Most importantly, this result affirms that the Josephson effect is general for any type of spontaneous symmetry breaking, depending only on ground state order parameters. 

%%%%%%%%%%%%%%%%%%%%%%%%%%%%%

\textit{AC Josephson effect.}
Compared to the DC effect, the AC Josephson effect is even more ubiquitous. As soon as one regards the order parameters $\mathbf{m}_\mathrm{L}$ and $\mathbf{m}_\mathrm{R}$ as 
quantum fields
with their own dynamics, it is obvious that there will be an oscillation $\sim \te^{\ti \mu t}$ due to an imposed energy difference $\propto \mu$ between left and right systems. Indeed, the AC Josephson effect does not even require superconductivity and has been suggested to occur in normal metals~\cite{Gaury15}. In the general setting, it follows from adding an asymmetric chemical potential:
\begin{equation}\label{eq:bias Hamiltonian}
 H_\mu = \mu ( Q^\mathrm{r}_\mathrm{R} -Q^\mathrm{r}_\mathrm{L}).
\end{equation}
This term clearly commutes with the symmetry generators $Q^\mathrm{r}_\mathrm{L}$, $Q^\mathrm{r}_\mathrm{R}$ separately and does not modify the DC effect. The order parameter operators $\mathcal{O}_\mathrm{L}$, $\mathcal{O}_\mathrm{R}$ in most cases do not commute with the Hamiltonian $H_0$, which naively implies a time dependence. This is associated with the Anderson tower of states and leads to an extremely slow oscillation of the order parameter expectation value with time scale that grows with the volume, and which vanishes in the thermodynamic limit~\cite{Anderson52}. We will neglect this trivial time dependence, and calculate:
\begin{align}\label{eq:AC Josephson effect}
  \partial_t\mathbf{m}_{\mathrm{L},\mathrm{R}} &=  \langle \partial_t \mathcal{O}_{\mathrm{L},\mathrm{R}} \rangle 
 = \mp \frac{\ti \mu}{\hbar}\langle [Q^\mathrm{r}_{\mathrm{L},\mathrm{R}},\mathcal{O}_{\mathrm{L},\mathrm{R}}]  \rangle   \notag\\
  &= \pm \frac{\ti \mu}{\hbar}  T^\mathrm{r} \mathbf{m}_{\mathrm{L},\mathrm{R}}. 
\end{align}
If we take the order parameters to be $\mathbf{m}_{\mathrm{L},\mathrm{R}} = m_{\mathrm{L},\mathrm{R}} \te^{\ti \alpha_{\mathrm{L},\mathrm{R}} T^\mathrm{r}} \mathbf{m}_0$, neglecting higher-order effects by setting $\partial_t \mathbf{m}_0 = 0$, $\partial_t m_{\mathrm{L},\mathrm{R}} = 0$, we can derive the equations of motion for the phases:
\begin{align}\label{eq:AC Josephson phase}
 \partial_t \alpha_{\mathrm{L},\mathrm{R}} &= \mp \frac{\mu}{\hbar}.
\end{align}
Integrating this equation gives the time-dependent phase difference $\alpha_{\mathrm{L},\mathrm{R}}(t) =  \mp \frac{\mu}{\hbar}t + \alpha_{\mathrm{L},\mathrm{R}}^0$. Denoting $\alpha^0 = \alpha_\mathrm{R}^0 -\alpha_\mathrm{L}^0$, the DC and AC effects can be combined as: 
\begin{equation}
 I = \frac{K}{\hbar}m_\mathrm{L}m_\mathrm{R}  \mathbf{m}_0^\dagger T^\mathrm{r} \sin \left( ( \frac{2\mu}{\hbar}t + \alpha^0) T^\mathrm{r} \right) \mathbf{m}_0.
\end{equation}

%%%%%%%%%%%%%%%%%%%%%%%%%%%%%%%%%%

\textit{Josephson energy.}
There is a potential energy associated with two systems that break the symmetry differently. We can simply calculate
\begin{align}
 \langle H_{K'}  + H_K \rangle 
 &= K (m_\mathrm{L} m_\mathrm{R} - \mathbf{m}_\mathrm{L} \cdot  \mathbf{m}_\mathrm{R}) \notag\\
 &= K m_\mathrm{L} m_\mathrm{R} (1 - \cos \alpha),
\end{align}
where for convenience we have chosen a real vector representation, and $\alpha$ is the angle between the two vectors. 
It is a remarkable fact that two initially causally disconnected systems that form order independently due to uncontrollable symmetry-breaking  dynamics, contain potential energy when brought close together.

The same result can be found in the standard way, by calculating the work needed to increase the phase difference from $\alpha(t=0) = 0$ to $\alpha(t) = \alpha$ as
$ U = \int_0^t \td t\;  I \mu$.
From Eqs.~\eqref{eq:boundary current}, \eqref{eq:bias Hamiltonian} one immediately sees that this indeed has units of energy. Substituting Eqs.~\eqref{eq:generalized Josephson effect commute} and Eq.~\eqref{eq:AC Josephson phase}:
\begin{align}\label{eq:Josephson energy}
  U &= \int_0^t \td t\; K m_\mathrm{L}m_\mathrm{R} \mathbf{m}_0^\dagger T^\mathrm{r} \sin \left(\alpha(t) T^\mathrm{r} \right) \mathbf{m}_0 \partial_t \alpha(t) \notag\\ 
  &= - K m_\mathrm{L}m_\mathrm{R} \mathbf{m}_0^\dagger \left( \int_0^t \td t\; \partial_t  \cos \left(\alpha(t) T^\mathrm{r} \right) \right) \mathbf{m}_0 \notag\\
  &=  Km_\mathrm{L}m_\mathrm{R}  \mathbf{m}_0^\dagger \left( 1 -  \cos (\alpha T^\mathrm{r}) \right)\mathbf{m}_0.
\end{align}

%%%%%%%%%%%%%%%%%%%%%%%%%%%%%

We shall now demonstrate these effects at the hand of several examples.

\textit{Superconductor/superfluid.}
Superconductors and superfluids are described by a complex scalar order parameter operator field $\psi$, with commutation relation $[\psi(x) ,\psi^\dagger(x')] = \delta(x -x')$ and expectation value $\langle \psi \rangle = \psi$. The Lagrangian is invariant under global $U(1)$ rotations $\psi \to \te^{ - \ti \alpha} \psi$, so that $T = - 1$. The Noether charge is $Q = \int \td^3 x\; \psi^\dagger \psi$. 
The coupling term Eq.~\eqref{eq:Josephson coupling} is:
\begin{equation}
 H_K = -\frac{K}{2} \left(\psi_\mathrm{L}^\dagger \psi_\mathrm{R} + \psi_\mathrm{R}^\dagger \psi_\mathrm{L} \right),
\end{equation}
from which the Feynman equations for the Josephson effect can be derived~\cite{Feynman65}. Taking $\psi_\mathrm{L} = \lvert \psi_\mathrm{L} \rvert $ and $\psi_\mathrm{R} = \te^{-\ti \alpha} \lvert \psi_\mathrm{R} \rvert $, Eqs.~\eqref{eq:generalized Josephson effect commute}, \eqref{eq:boundary current} give:
\begin{equation}
 I = \frac{K}{\hbar} \lvert \psi_\mathrm{L} \rvert \lvert \psi_\mathrm{R} \rvert  \sin \alpha,
\end{equation}
which is the standard result for the DC Josephson current. The AC Josephson effect Eq.~\eqref{eq:AC Josephson phase} is:
\begin{equation}
 \partial_t \alpha = -\frac{2\mu}{\hbar} = \frac{2 e V}{\hbar},
\end{equation}
where we have set $\mu = -e V$ with $e$ the electron charge and $V$ the electrostatic potential, to reproduce the standard AC Josephson equation. Similarly the Josephson energy Eq.~\eqref{eq:Josephson energy} is:
\begin{equation}
 U = K \lvert \psi_\mathrm{L} \rvert \lvert \psi_\mathrm{R} \rvert  \left( 1 -  \cos \alpha \right).
\end{equation}

%%%%%%%%%%%%%%%%%%%%%%%%%%%%%

\textit{Heisenberg magnet.}
The Heisenberg Hamiltonian on a bipartite lattice with sites $i$ is
\begin{equation}
 H_0 = \frac{J}{2} \sum_{ i \delta } \mathbf{S}_i \cdot \mathbf{S}_{i+\delta}.
\end{equation}
Here $\mathbf{S}_i$ are the $SU(2)$ spin operators on site $i$ with commutation relations $[S^a_i,S^b_j] = \ti \epsilon_{abc} S^c_i \delta_{ij}$, $\delta$ runs over all elementary lattice vectors and $J$ is the coupling energy. The system prefers a ferromagnetic configuration when $J<0$ and antiferromagnetic (N\'eel) configuration when $J>0$. This Hamiltonian is invariant under global $SU(2)$ spin rotations with generators $S^a = \sum_i S^a_i$. The Hamiltonian is not in canonical form and we cannot perform the Legendre transform to a Lagrangian. Still the lattice Noether current $j^a_{i,i+\delta}$, which is the current that flows through the Josephson junction, can be obtained directly from the equation of motion:
\begin{equation}
 \partial_t S^a_i = - \sum_\delta \epsilon_{abc} (S^b_i S^c_{i+\delta} + S^c_{i}S^b_{i+\delta}) \equiv - \sum_\delta j^a_{i,i+\delta}.
\end{equation}
If $\mathbf{m}(x)$ is a classical vector corresponding to the magnetic moment at $x$, then this equation is the discretized, quantum version of (cf. Ref.~\cite{Ruckriegel17}):
\begin{equation}
 \partial_t \mathbf{m} = - \partial_m ( \mathbf{m} \times \partial_m \mathbf{m}) \equiv - \nabla \cdot \mathbf{j}.
\end{equation}
A ferromagnet has magnetization $\mathbf{M}_i = S^z_i$ as order parameter, while for an antiferromagnet it is  staggered magnetization $\mathbf{M}_i = (-1)^i S^z_i$, which each break two spin-rotations $S^x$, $S^y$ while rotations $S^z$ around the (staggered) magnetization direction are still symmetries. The symmetry generators $S^a$ act on (staggered) magnetization vectors $\mathbf{m} = \langle \mathbf{M} \rangle$ in the vector representation $(T^a)_{ij} = \ti \epsilon_{aij}$. Without loss of generality we take the configuration where  $\mathbf{m}_\mathrm{L} = M_\mathrm{L} \mathbf{m}_0$ and $\mathbf{m}_\mathrm{R} =  M_\mathrm{R} \te^{\ti \alpha T^x} \mathbf{m}_0 $, with $\mathbf{m}_0$ the unit vector $(0,0,1)$ and $M_\mathrm{L}$, $M_\mathrm{R}$ the magnitude of the total (staggered) magnetization of the left and right systems. The coupling Hamiltonian Eq.~\eqref{eq:Josephson coupling} is:
\begin{equation}
 H_K = -K \mathbf{M}_\mathrm{L} \cdot \mathbf{M}_\mathrm{R}.
\end{equation}
because the order parameter operators are Hermitian. Then we find the DC spin Josephson current Eq.~\eqref{eq:generalized Josephson effect}:
\begin{align}
 \langle \partial_t S^x \rangle &= - \frac{K}{\hbar} M_\mathrm{L} M_\mathrm{R} \mathbf{m}_0^\dagger T^x \sin(\alpha T^x) \mathbf{m}_0 \notag\\
 &= - \frac{K}{\hbar} M_\mathrm{L} M_\mathrm{R}\sin \alpha.
\end{align}
One can also explicitly verify that for these order parameters $\langle \partial_t S^y \rangle = \langle \partial_t S^z \rangle = 0$. This result agrees with the usual form $\partial_t \mathbf{m} = \mathbf{m}_\mathrm{L} \times \mathbf{m}_\mathrm{R}$ for ferromagnets~\cite{Lee03,Chen14,Ruckriegel17}. For the antiferromagnet the equations are usually given in terms of the staggered magnetization vector~\cite{Chasse10,Moor12,Liu16}, but here we have shown that is rather the spin current that is flowing. The Josephson energy of this configuration is
\begin{equation}
 U = K M_\mathrm{L} M_\mathrm{R}\left( 1 -  \cos \alpha \right).
\end{equation}

The AC Josephson effect follows from the Hamiltonian $H_\mu = \bm{\mu} \cdot (\mathbf{S}_\mathrm{L} - \mathbf{S}_\mathrm{R})$ where $\bm{\mu}$ is proportional to an external magnetic field imposed in opposite directions for left and right systems. While Eq.~\eqref{eq:AC Josephson effect} is valid, it is more insightful to simply derive:
\begin{align}
 \partial_t \mathbf{m}_\mathrm{L} &= -\bm{\mu} \times \mathbf{m}_\mathrm{L}, &
 \partial_t \mathbf{m}_\mathrm{R} &= \bm{\mu} \times \mathbf{m}_\mathrm{R}.
\end{align}
which {\em mutatis mutandis} agrees with Ref.~\cite{Chasse10}. 

%%%%%%%%%%%%%%%%%%%%%%%%%%%%%

\textit{Helium-3.} 
A helium-3 superfluid has triplet pairing and an anisotropic order parameter transforming under rotations ($L$), spin rotations ($S$) and global phase rotations, with a total symmetry group $G = SO(3)_L\times SO(3)_S \times U(1)$, with seven generators spanning the Lie algebra. There are many symmetry breaking patterns~\cite{Vollhardt90}, but we will here look at the B-phase where the spin is locked to angular momentum, with residual symmetry group $H = SO(3)_{L + S}$. There are four broken generators; the order parameter space is isomorphic to $G/H \simeq SO(3) \times U(1)$. The $U(1)$ part follows the pattern of the ordinary superfluid above, so we focus on the $SO(3)$ subspace. The order parameter is a fixed orthogonal $3 \times 3$ matrix $A_{\mu j}$, where $\mu$ transforms under $\mathbf{S}$ and $j$ under $\mathbf{L}$, denoting the rotation from the position to the spin coordinate frame~\cite{Vollhardt90}.
The three broken generators $L^a-S^a$ are represented by the matrices ${T^a}_{ij} = \ti \epsilon_{aij}$ acting on $A$. Suppose that $\mathbf{m}_\mathrm{L} = m_\mathrm{L}A$ and $\mathbf{m}_\mathrm{R} = m_\mathrm{R}\te^{\ti \alpha T^\mathrm{r}}A$. The DC Josephson current Eq.~\eqref{eq:generalized Josephson effect commute} is:
\begin{align}
I^\mathrm{r} &= \frac{K}{\hbar}  m_\mathrm{L} m_\mathrm{R} \mathrm{Tr} \left(A^\mathrm{T} T^\mathrm{r} \sin(\alpha T^{\mathrm{r}} )A \right) %\notag\\
%&
= \frac{K}{\hbar}  m_\mathrm{L} m_\mathrm{R} 2 \sin\alpha, \notag
\end{align}
where we used the cyclic property of the matrix trace $\mathrm{Tr}$ and $A A^\mathrm{T} = 1$. The current $I^a$ for $a \neq \mathrm{r}$ vanishes.

%%%%%%%%%%%%%%%%%%%%%%%%%%%%%

\begin{figure}[t]
 \includegraphics[width=.85\columnwidth]{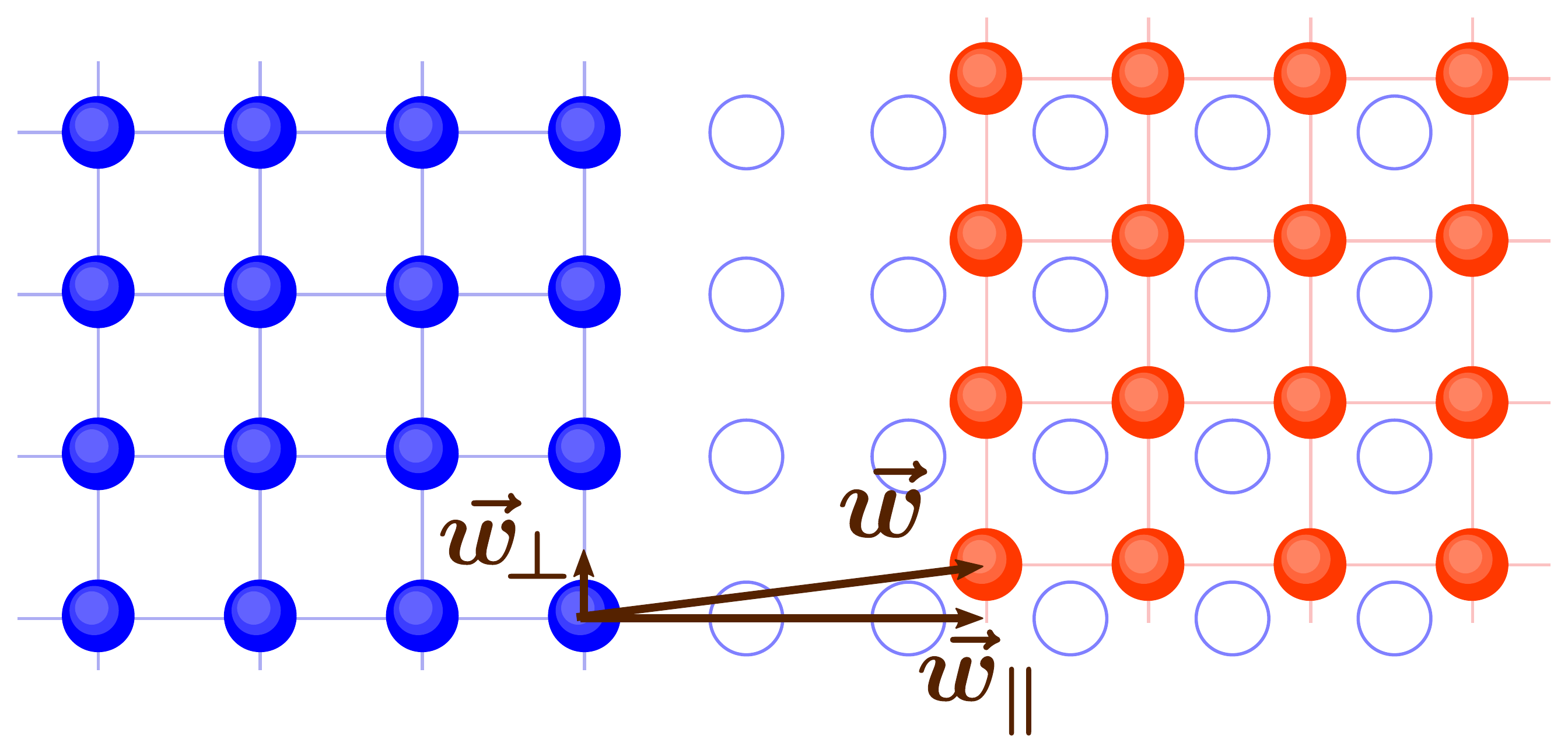}
 \caption{The crystalline DC Josephson effect occurs when left (blue) and right (red) crystalline solids with identical lattice
are displaced by a vector $\vec{w}$ that is is not a lattice vector. The empty blue dots indicate the order that is preferred by the left crystal.}
 \label{fig:crystals}
\end{figure}

\textit{Crystalline solid.}
The most interesting prediction that follows from Eq.~\eqref{eq:generalized Josephson effect} is a Josephson effect for $d$-dimensional solids. It is necessary to carefully determine the order parameter capturing the translational symmetry breaking $\mathbb{R}^d \to \mathbb{Z}^d$ to discrete lattice translations. A medium is described by a field $\phi_i(\mathbf{x})$
denoting the comoving coordinates of the 
the volume element at $\mathbf{x}$. The action is invariant under internal shifts $\phi_i(\mathbf{x}) \to \phi_i(\mathbf{x}) + \alpha^a \delta_{ia}$, for which the symmetry generator is $Q^a = \Pi_a  = \int_V \pi_a$, and $\pi_i(\mathbf{x})$ the canonical momentum conjugate to $\phi_i(\mathbf{x})$ with $[\phi_i(\mathbf{x}),\pi_j(\mathbf{x}')] = \ti \delta_{ij} \delta(\mathbf{x} - \mathbf{x'})$. For a homogeneous medium we have $\langle \phi_i(\mathbf{x})\rangle = x_i$, and this suffices to break the symmetry~\cite{Nicolis15}. For a solid, we need to preserve lattice translations, and the order parameter should take values in $G/H \simeq U(1)^{\otimes d}$. We write the order parameter as $\mathcal{O}(\mathbf{x}) = \otimes_{i=1}^d  \mathcal{O}_i(\mathbf{x})$, with 
\begin{equation}
 \mathcal{O}_i(\mathbf{x}) = \te^{ 2\pi \ti \phi_i(\mathbf{x}) / c_i} \rho(\mathbf{x}),
\end{equation}
where $c_i$ is the lattice constant in direction $i$ and $\rho(\mathbf{x})$ the density. This quantity is invariant under lattice translations $\phi_i \to \phi_i + n c_i, n \in \mathbb{Z}$. We have $[Q^a, \mathcal{O}_i] = \frac{2\pi}{c_i} \delta_{ai} \mathcal{O}_i$, represented by $T^a_{ij} = \frac{2\pi}{c_a} \delta_{ai} \delta_{ij} $.
For a lattice with primitive unit cells, the density is $\rho(\mathbf{x}) = \prod_{i=1}^d \sum_{n_i \in \mathbb{Z}} \delta\left(x_i - (n_i + \varphi_i)c_i \right)$. Here $\varphi_i \in[0,1)$ is the position of the origin of the unit cell with respect to the coordinate system $x$,  in units of $c_i$, modulo lattice translations. The order parameter expectation value is, averaging over one unit cell: $\langle \mathcal{O}_i(\mathbf{x}) \rangle = \te^{\ti 2\pi \varphi_i}$,

For the DC Josephson effect, we regard two crystals with perfect surface along a crystal direction, separated by a short distance $w_\parallel$, and relatively displaced along the surface by $w_\perp$, see Fig.~\ref{fig:crystals}. The Josephson effect occurs for each lattice direction $a$ when $w_a$ is not an exact multiple of the lattice constant $c_a$. (For the longitudinal effect, it is necessary that the medium in between can support the field $\phi_i(\mathbf{x})$ over the distance $w_\parallel$.) This leads to a displacement (phonon) current from the left to the right system:
\begin{equation}
 I^a = \frac{K}{\hbar} \frac{2\pi}{c_a} \sin \Delta_a, \qquad \Delta_a = 2\pi\frac{w_a}{c_a}.
\end{equation}
The Josephson energy is $U = \sum_a K ( 1 - \cos \Delta_a)$. In contrast with Josephson effects related to broken internal symmetries, the energy depends periodically on displacement $\mathbf{w}$ between the two crystals (on top of a possible $w_\parallel$-dependence of the coupling $K$). Then there is an associated force $F_a = - \partial_{w_a} U = K \frac{2\pi}{c_a} \sin \Delta_a$ (neglecting $\partial_{w_a} K$). This Josephson force is purely due to the symmetry breaking and adds to or competes with for instance Casimir/Van der Waals forces.

%%%%%%%%%%%%%%%%%%%%%%%%%%%%%

\textit{Outlook.}
We have shown that the Josephson effect is intrinsic to the phenomenon of spontaneous symmetry breaking.  Other generalized phenomena similarly appear, such as the Shapiro effect when the applied current is periodic. This paves the way to explore hitherto unaddressed Josephson currents. It should be kept in mind that it is in general necessary to apply an external current in order to detect the Josephson current, and this may not be straightforward in some settings. Atomic condensates may form a source of inspiration~\cite{Sukhatme2001}. For the effect in crystalline solids, measurement of the force $\mathbf{F} = - \nabla U$ looks most promising. 
Its magnitude is difficult to estimate in full generality. The tunneling of phonons should fall off exponentially with $w_\parallel$, with a length scale $\xi$ that is the healing length for crystalline order. This should be related to the core size of dislocations~\cite{Peierls40}.
%%%%%%%%%%%%%%%%%%%%%%%%%%%%%

\textit{Acknowledgments.}
I thank Jasper van Wezel and Louk Rademaker for collaboration on a related project that inspired this work and for many useful discussions, and Naoki Yamamoto and Antonino Flachi for careful reading of the manuscript. This work is supported by the MEXT-Supported Program for the Strategic Research Foundation at Private Universities ``Topological Science'' (Grant No. S1511006) and by JSPS Grant-in-Aid for Early-Career Scientists (Grant No. 18K13502).

\bibliography{GJEreferences}

\end{document}